# SECOND-ORDER ADJOINT SENSITIVITY ANALYSIS PROCEDURE (SO-ASAP) FOR COMPUTING EXACTLY AND EFFICIENTLY FIRST- AND SECOND-ORDER SENSITIVITIES IN LARGE-SCALE LINEAR SYSTEMS: I. COMPUTATIONAL METHODOLOGY

Dan G. Cacuci

Department of Mechanical Engineering, University of South Carolina
E-mail: cacuci@cec.sc.edu

Corresponding author:

Department of Mechanical Engineering, University of South Carolina

300 Main Street, Columbia, SC 29208, USA

Email: cacuci@cec.sc.edu; Phone: (919) 909 9624; Submitted to JCP: August 14, 2014

**ABSTRACT**

This work presents the *second-order forward and adjoint sensitivity analysis procedures* (*SO-FSAP* and *SO-ASAP*) for computing exactly and efficiently the second-order functional derivatives of physical (engineering, biological, etc.) system responses (i.e., "system performance parameters") to the system's model parameters. The definition of "system parameters" used in this work includes all computational input data, correlations, initial and/or boundary conditions, etc. For a physical system comprising $N_\alpha$ parameters and $N_r$ responses, we note that the *SO-FSAP* requires a total of $\left(N_\alpha^2/2 + 3N_\alpha/2\right)$ large-scale computations for obtaining all of the first- and second-order sensitivities, for all $N_r$ system responses. On the other hand, the *SO-ASAP* requires a total of $(2N_\alpha + 1)$ large-scale computations for obtaining all of the first- and second-order sensitivities, for one functional-type system responses. Therefore, the *SO-FSAP* should be used when $N_r \gg N_\alpha$, while the *SO-ASAP* should be used when $N_\alpha \gg N_r$. The original *SO-ASAP* presented in this work should enable the hitherto very difficult, if not intractable, *exact* computation of *all* of the second-order response sensitivities (i.e., functional Gateaux-derivatives) for large-systems involving many parameters, as usually encountered in practice. Very importantly, the implementation of the *SO-ASAP* requires very little additional effort beyond the construction of the adjoint sensitivity system needed for computing the first-order sensitivities.

# 1. INTRODUCTION

The purpose of this paper is to present a new method for computing, exactly and efficiently, *second-order functional derivatives of system responses* (i.e., "system performance parameters" in physical, engineering, biological systems) *to the system's model parameters*, which are considered in this work in the most comprehensive sense, i.e., including all input data, correlations, initial and/or boundary conditions, etc. This new method builds on the first-order adjoint sensitivity analysis procedure (*ASAP*) for nonlinear systems introduced ([1], [2]) and developed ([3]-[5]) by Cacuci; see also [6]. The aims and means of sensitivity theory/analysis are occasionally confused with the aims and means of optimization theory; such confusions usually arise because of some shared terminology, and because some of the mathematical derivations underlying sensitivity theory appear (superficially) to be similar to those underlying optimization theory. To delineate clearly the aims this work, it is therefore useful to recall briefly the terminology and aims of sensitivity theory/analysis, while highlighting the distinctions to optimization theory.

In general, a physical system and/or the result of an indirect experimental measurement is modeled mathematically in terms of:

(a) A system of linear and/or nonlinear equations that relate the system's independent variables and parameters to the system's state (i.e., dependent) variables;

(b) Probability distributions, moments thereof, inequality and/or equality constraints that define the range of variations of the system's parameters;

(c) One or several quantities, customarily referred to as system responses (or objective functions, or indices of performance), which are computed using the mathematical model.

Sensitivity theory/analysis aims at investigating the variation of system responses caused by variations in the system's parameters, while optimization theory aims at optimizing (minimizing or maximizing) the system responses of interest. Therefore, both sensitivity theory and optimization compute first- and occasionally second-order response derivatives with respect to the state functions and/or parameters. The *ASAP* uses the solution of certain adjoint equations, which we call the *adjoint sensitivity system*, for computing efficiently all of the first-order response sensitivities. Adjoint equations also appear in constrained optimization algorithms, particularly in those used in control theory and for data assimilation in the atmospheric and earth sciences (see, e.g., [7] and [8] for recent expositions), where the Lagrange multipliers used to append the constraints to the "objective" response functional are determined as the solutions of certain adjoint equations. However, in contrast to optimization theory, in which some or all of the first-order response derivatives are driven by the respective computational algorithm to zero, sensitivity theory evaluates first- (and occasionally higher-) order response derivatives with respect to model parameters at the parameters' and state functions' nominal values. Thus, the adjoint systems/equations found in constrained optimization and, respectively, sensitivity analysis serve different purposes and conceptually differ from each other. These important distinctions have been thoroughly analyzed [9] in the setting of both local and global optimization and, respectively, local and global sensitivity analysis. Since this paper deals exclusively with sensitivity theory, the vast literature dealing with the computation of response derivatives for the purposes of system response, or "cost functional," optimization will not be discussed further.

Response sensitivities to model parameters are needed in many activities, including:
(i) understanding the system by identifying and ranking the importance of model parameters in influencing the response under consideration;
(ii) determining the effects of parameter variations on the system's behavior;

(iii) improving the system design, possibly reducing conservatism and redundancy;

(iv) prioritizing possible improvements for the system under consideration;

(v) quantifying uncertainties in responses due to quantified parameter uncertainties (e.g., by using the method of "propagation of uncertainties");

(vi) performing "predictive modeling" (which includes data assimilation and model calibration) for the purpose of obtaining best-estimate predicted results with reduced predicted uncertainties.

First-order response sensitivities can be computed by either statistical or deterministic methods; the most popular of these methods were reviewed in [5]. In practice, sensitivities cannot be computed exactly using statistical methods; this can only be done (for obtaining exact analytical expressions and/or values computed to machine accuracy, in principle) using deterministic methods. For a system comprising $N_\alpha$ parameters and $N_r$ responses, forward deterministic methods require $O(N_\alpha)$ large-scale forward model computations while the *ASAP* requires $O(N_r)$ large-scale adjoint model computations. Therefore, when the number of model responses exceeds the number of model responses of interest, forward methods (e.g., difference methods using re-computations, the forward sensitivity analysis procedure) are computationally most efficient for computing sensitivities. On the contrary, when the number of model parameters exceeds the number of responses of interest, which is invariably the case in practice for large-scale nonlinear problems, the most efficient method for computing sensitivities exactly is the adjoint sensitivity analysis procedure (*ASAP*).

The earliest attempts at computing second-order response derivatives to model parameters for large-scale systems appear to have been made in the field of nuclear reactor physics, as highlighted in [10], [11], which computed functional-type response sensitivities to the thousand of neutron and photon cross sections (system parameters) within the context of the linear six-dimensional neutron and photon Boltzmann transport equation. The best achievements of these early works was a selective and approximate computation of a few second-order sensitivities, obtained essentially by finite-differencing first-order sensitivities computed using the adjoint neutron/photon transport equation. The current state-of-the-art methods in the geophysical sciences (see, e.g., [12] and references therein), for example, are able to compute first-order sensitivities of a user-defined "cost functional" that describes the "misfit" between computations and measurements, along with the product between a user-defined vector and the Hessian matrix of this cost functional with respect to the state variables (not the system parameters!). The current methods still require $O(N_\alpha^2)$ large-scale computations for computing all of the second-order response sensitivities (functional derivatives) to all $N_\alpha$ model parameters, which severely strains computational resources when $N_\alpha$ is large. We are not aware of any successful achievements to compute exactly the complete matrix of second-order sensitivities (functional derivatives) of a system response to all of the model/system parameters, using current methodologies.

This new *SO-ASAP* computational method presented in this work requires only at most $O(N_\alpha)$, rather than in $O(N_\alpha^2)$, large-scale computations for quantifying all of the (mixed) second-oder response sensitivities for all of the model parameters. As already mentioned, the foundation for the *SO-ASAP* is provided by the first-order adjoint sensitivity analysis procedure (*ASAP*), which was introduced in [1] and [2], and will be recalled in Section 2 of this work. Building on the *ASAP*, Section 3 presents the *SO-ASAP*. Section 4 concludes this work, by highlighting the significance of the SO-ASAP and the directions that we are currently pursuing for further generalizing the SO-ASAP, and prepares the stage for the comprehensive paradigm application to a particle diffusion problem, which will be presented in the sequel, PART II, to this work.

## 2. BACKGROUND: THE FIRST-ORDER *ASAP* FOR LARGE-SCALE LINEAR SYSTEMS

Consider that the physical system is represented mathematically by means of $K_u$ coupled linear operator equations of the form

$$\sum_{j=1}^{K_u} L_{ij}(\boldsymbol{\alpha}) u_j(\mathbf{x}) = Q_i(\boldsymbol{\alpha}), \ i = 1, \ldots, K_u, \quad \mathbf{x} \in \Omega_x, \tag{1.a}$$

which can be written in matrix form as

$$\mathbf{L}[\boldsymbol{\alpha}(\mathbf{x})]\mathbf{u}(\mathbf{x}) = \mathbf{Q}[\boldsymbol{\alpha}(\mathbf{x})], \quad \mathbf{x} \in \Omega_x, \tag{1.b}$$

where:

1. $\mathbf{x} = (x_1, \ldots, x_{J_x})$ denotes the $J_x$-dimensional phase-space position vector for the primary system; $\mathbf{x} \in \Omega_x \subset \mathbb{R}^{J_x}$, where $\Omega_x$ is a subset of the $J_x$-dimensional real vector space $\mathbb{R}^{J_x}$;

2. $\mathbf{u}(\mathbf{x}) = [u_1(\mathbf{x}), \ldots, u_{K_u}(\mathbf{x})]$ denotes a $K_u$-dimensional column vector whose components are the system's dependent (i.e., state) variables; $\mathbf{u}(\mathbf{x}) \in \mathscr{E}_u$, where $\mathscr{E}_u$ is a normed linear space over the scalar field $\mathscr{F}$ of real numbers;

3. $\boldsymbol{\alpha}(\mathbf{x}) = [\alpha_1(\mathbf{x}), \ldots, \alpha_{N_\alpha}(\mathbf{x})]$ denotes an $N_\alpha$-dimensional column vector whose components are the system's parameters; $\boldsymbol{\alpha} \in \mathscr{E}_\alpha$, where $\mathscr{E}_\alpha$ is also a normed linear space;

4. $\mathbf{Q}[\boldsymbol{\alpha}(\mathbf{x})] = [Q_1(\boldsymbol{\alpha}), \ldots, Q_{K_u}(\boldsymbol{\alpha})]$ denotes a $K_u$-dimensional column vector whose elements represent inhomogeneous source terms that depend either linearly or nonlinearly on $\boldsymbol{\alpha}$; $\mathbf{Q} \in \mathscr{E}_Q$, where $\mathscr{E}_Q$ is also a normed linear space; the components of $\mathbf{Q}$ may be operators, rather than just functions, acting on $\boldsymbol{\alpha}(\mathbf{x})$ and $\mathbf{x}$;

5. $\mathbf{L} \equiv [L_1(\boldsymbol{\alpha}), \ldots, L_{K_u}(\boldsymbol{\alpha})]$ denotes a $K_u$-component column vector whose components are *operators* (including differential, difference, integral, distributions, and/or infinite matrices) acting *linearly* on $\mathbf{u}$ and *nonlinearly* on $\boldsymbol{\alpha}$. For notational convenience, all vectors in this work are considered to be column vectors; transposition will be indicated by a dagger $(\dagger)$.

In view of the definitions given above, $\mathbf{L}$ represents the mapping $\mathbf{L}:\mathscr{D}\subset\mathscr{E}\to\mathscr{E}_Q$, where $\mathscr{D}=\mathscr{D}_u\times\mathscr{D}_\alpha$, $\mathscr{D}_u\subset\mathscr{E}_u$, $\mathscr{D}_\alpha\subset\mathscr{E}_\alpha$, and $\mathscr{E}=\mathscr{E}_u\times\mathscr{E}_\alpha$. Note that an arbitrary element $\mathbf{e}\in\mathscr{E}$ is of the form $\mathbf{e}=(\mathbf{u},\boldsymbol{\alpha})$. If differential operators appear in Eq. (1), then a corresponding set of boundary and/or initial conditions (which are essential to define $\mathscr{D}$) must also be given. Since we consider here only systems that are linear in the state function $\mathbf{u}(\mathbf{x})\in\mathscr{E}_u$, the accompanying boundary and/or initial condition s must also be linear in $\mathbf{u}(\mathbf{x})$, so they can therefore be represented in operator form as

$$\left[\mathbf{B}(\boldsymbol{\alpha})\mathbf{u} - \mathbf{A}(\boldsymbol{\alpha})\right]_{\partial\Omega_x} = \mathbf{0}, \quad \mathbf{x}\in\partial\Omega_x, \tag{2}$$

where $\partial\Omega_x$ denotes the boundary of $\Omega_x$ while $\mathbf{A}$ and $\mathbf{B}$ denote operators that act *nonlinearly* on the model parameters $\boldsymbol{\alpha}$, but $\mathbf{B}(\boldsymbol{\alpha})\mathbf{u}$ acts *linearly* on $\mathbf{u}$ but *nonlinearly* on $\boldsymbol{\alpha}$.

The vector-valued function $\mathbf{u}(\mathbf{x})$ is considered to be the unique nontrivial solution of the physical problem described by Eqs. (1) and (2). The system response (i.e., result of interest), associated with the problem modeled by Eqs. (1) and (2) will be denoted here as $R(\mathbf{u},\boldsymbol{\alpha})$; in this work, $R(\mathbf{u},\boldsymbol{\alpha})$ is considered to be a real-valued *nonlinear functional* of $(\mathbf{u},\boldsymbol{\alpha})$, which can be generally represented in operator form as

$$R(\mathbf{u},\boldsymbol{\alpha}):\mathscr{D}_R\subset\mathscr{E}\to\mathscr{F}, \tag{3}$$

where $\mathscr{F}$ denotes the field of real scalars.

The nominal parameter values $\boldsymbol{\alpha}^0(\mathbf{x})$ are used in Eqs. (1) and (2) to obtain the nominal solution $\mathbf{u}^0(\mathbf{x})$ by solving these equations; mathematically, therefore, the nominal value $\mathbf{u}^0(\mathbf{x})$ of the state-function is obtained by solving

$$\mathbf{L}(\boldsymbol{\alpha}^0)\mathbf{u}^0 = \mathbf{Q}(\boldsymbol{\alpha}^0), \quad \mathbf{x} \in \Omega_x, \tag{4}$$

$$\mathbf{B}(\boldsymbol{\alpha}^0)\mathbf{u}^0 = \mathbf{A}(\boldsymbol{\alpha}^0), \quad \mathbf{x} \in \partial\Omega_x. \tag{5}$$

Equations (4) and (5) represent the "*base-case*" or *nominal* state of the physical system. After solving Eqs. (4) and (5), the *nominal solution*, $\mathbf{u}^0(\mathbf{x})$, thus obtained is used to obtain the *nominal value* $R(\mathbf{e}^0)$ of the response $R(\mathbf{e})$, using the nominal values $\mathbf{e}^0 = (\mathbf{u}^0, \boldsymbol{\alpha}^0) \in \mathscr{E}$ of the model's state function and parameters.

Next, we consider (a vector of) arbitrary variations $\mathbf{h} \equiv (\mathbf{h}_u, \mathbf{h}_\alpha) \in \mathscr{E} = \mathscr{E}_u \times \mathscr{E}_\alpha$, with $\mathbf{h}_u \equiv (\delta u_1, \ldots, \delta u_{K_u}) \in \mathscr{E}_u$ and $\mathbf{h}_\alpha \equiv (\delta \alpha_1, \ldots, \delta \alpha_{N_\alpha}) \in \mathscr{E}_\alpha$, around $\mathbf{e}^0 = (\mathbf{u}^0, \boldsymbol{\alpha}^0) \in \mathscr{E}$. The variation (sensitivity) of the response $R$ to variations $\mathbf{h}$ in the system parameters is given by the Gâteaux- (G-)differential $\delta R(\mathbf{e}^0; \mathbf{h})$ of the response $R(\mathbf{e})$ at $\mathbf{e}^0 = (\mathbf{u}^0, \boldsymbol{\alpha}^0)$ with increment $\mathbf{h}$, which is defined as

$$\delta R(\mathbf{e}^0; \mathbf{h}) \equiv \left\{ \frac{d}{d\varepsilon}\left[R(\mathbf{e}^0 + \varepsilon \mathbf{h})\right] \right\}_{\varepsilon=0} = \lim_{\varepsilon \to 0} \frac{R(\mathbf{e}^0 + \varepsilon \mathbf{h}) - R(\mathbf{e}^0)}{\varepsilon}, \tag{6}$$

for $\varepsilon \in \mathscr{F}$, and all (i.e., arbitrary) vectors $\mathbf{h} \in \mathscr{E}$. When the response $R(\mathbf{e})$ is functional of the form $R: \mathscr{D}_R \to \mathscr{F}$, the sensitivity $\delta R(\mathbf{e}^0; \mathbf{h})$ is also an operator, defined on the same domain, and with the same range as $R(\mathbf{e})$. The G-differential $\delta R(\mathbf{e}^0; \mathbf{h})$ is related to the total variation $\left[R(\mathbf{e}^0 + \varepsilon \mathbf{h}) - R(\mathbf{e}^0)\right]$ of $R$ at $\mathbf{e}^0$ through the relation

$$R(\mathbf{e}^0 + \varepsilon \mathbf{h}) - R(\mathbf{e}^0) = \delta R(\mathbf{e}^0; \mathbf{h}) + \Delta(\mathbf{h}), \quad \text{with} \quad \lim_{\varepsilon \to 0}\left[\Delta(\varepsilon \mathbf{h})\right]/\varepsilon = 0. \tag{7}$$

As discussed in [1] and [2], *the most general definition of the first-order sensitivity of a response to variations in the model parameter is the G-differential* $\delta R(\mathbf{e}^0; \mathbf{h})$ defined in Eq. (6). Since the system's state vector $\mathbf{u}$ and parameters $\boldsymbol{\alpha}$ are related to each other through Eqs. (1) and (2), it follows that $\mathbf{h}_u$ and $\mathbf{h}_\alpha$ are also related to each other. Therefore, the sensitivity $\delta R(\mathbf{e}^0;\mathbf{h})$ of $R(\mathbf{e})$ at $\mathbf{e}^0$ can only be evaluated after determining the vector of variations $\mathbf{h}_u$ in terms of the vector of parameter variations $\mathbf{h}_\alpha$. The first-order relationship between $\mathbf{h}_u$ and $\mathbf{h}_\alpha$ is determined by taking the G-differentials of Eqs. (1) and (2). Thus, taking the G-differential at $\mathbf{e}^0$ of Eq. (1) yields

$$\sum_{j=1}^{K_u} L_{ij}(\boldsymbol{\alpha}^0)\delta u_j(\mathbf{x}) = \sum_{k=1}^{N_\alpha}\left[\frac{\partial Q_i(\boldsymbol{\alpha}^0)}{\partial \alpha_k} - \sum_{j=1}^{K_u}\frac{\partial L_{ij}(\boldsymbol{\alpha}^0)}{\partial \alpha_k}u_j^0(\mathbf{x})\right]\delta\alpha_k,\ i=1,...,K_u,\ \mathbf{x}\in\Omega_x,$$

which can be written in matrix as

$$\mathbf{L}(\boldsymbol{\alpha}^0)\mathbf{h}_u = \left\langle \mathbf{D}_\alpha \mathbf{Q}(\boldsymbol{\alpha}^0) - \mathbf{D}_\alpha\left[\mathbf{L}(\boldsymbol{\alpha}^0)\mathbf{u}^0\right], \mathbf{h}_\alpha \right\rangle_\alpha,\quad \mathbf{x}\in\Omega_x, \tag{8}$$

where

$$\mathbf{D}_\alpha \mathbf{Q}(\boldsymbol{\alpha}) \equiv \begin{pmatrix} \dfrac{\partial Q_1}{\partial \alpha_1} & \cdots & \dfrac{\partial Q_1}{\partial \alpha_{N_\alpha}} \\ \vdots & \ddots & \vdots \\ \dfrac{\partial Q_{K_u}}{\partial \alpha_1} & \cdots & \dfrac{\partial Q_{K_u}}{\partial \alpha_{N_\alpha}} \end{pmatrix};\ \mathbf{D}_\alpha\left[\mathbf{L}(\boldsymbol{\alpha})\mathbf{u}\right] \equiv \begin{pmatrix} \dfrac{\partial\left[\sum_{j=1}^{K_u} L_{1,j}(\boldsymbol{\alpha})u_j\right]}{\partial\alpha_1} & \cdots & \dfrac{\partial\left[\sum_{j=1}^{K_u} L_{1,j}(\boldsymbol{\alpha})u_j\right]}{\partial\alpha_{N_\alpha}} \\ \vdots & \ddots & \vdots \\ \dfrac{\partial\left[\sum_{j=1}^{K_u} L_{K_u,j}(\boldsymbol{\alpha})u_j\right]}{\partial\alpha_1} & \cdots & \dfrac{\partial\left[\sum_{j=1}^{K_u} L_{K_u,j}(\boldsymbol{\alpha})u_j\right]}{\partial\alpha_{N_\alpha}} \end{pmatrix}.$$

Taking the G-differential at $\mathbf{e}^0$ of the boundary and initial conditions represented by Eq. (2) yields

$$\mathbf{B}(\boldsymbol{\alpha}^0)\mathbf{h}_u = \left\langle \mathbf{D}_\alpha \mathbf{A}(\boldsymbol{\alpha}^0) - \mathbf{D}_\alpha \mathbf{B}(\boldsymbol{\alpha}^0)\mathbf{u}^0, \mathbf{h}_\alpha \right\rangle_\alpha,\quad \mathbf{x}\in\partial\Omega_x. \tag{9}$$

Equations (8) and (9) represent the "*forward sensitivity equations*" (*FSE*), which are also occasionally called the "forward sensitivity model," the "forward variational model", or the "tangent linear model." For a given vector of parameter variations $\mathbf{h}_\alpha$ around $\boldsymbol{\alpha}^0$, the forward sensitivity system represented by Eqs. (8) and (9) is solved to obtain $\mathbf{h}_u$. Once $\mathbf{h}_u$ is available, it is in turn used in Eq. (6) to compute the sensitivity $\delta R(\mathbf{e}^0;\mathbf{h})$ of $R(\mathbf{e})$ at $\mathbf{e}^0$, for a given vector of parameter variations $\mathbf{h}_\alpha$. *The direct computation of the response sensitivity* $\delta R(\mathbf{e}^0;\mathbf{h})$ by using the ($\mathbf{h}_\alpha$-dependent) solution $\mathbf{h}_u$ of Eqs. (8) and (9) is called [1] *the Forward Sensitivity Analysis Procedure (FSAP)*. From the standpoint of computational costs and effort, the *FSAP* requires require $O(N_\alpha)$ large-scale forward computations; therefore the *FSAP* is advantageous to employ only if, in the problem under consideration, the number $N_r$ of responses of interest exceeds the number of system parameters and/or parameter variations of interest. This is rarely the case in practice, however, since most problems of practical interest are characterized by many parameters (i.e., **α** has many components) and comparatively few responses. In such situations, it is not economical to employ the *FSAP* since it becomes prohibitively expensive to solve repeatedly the $\mathbf{h}_\alpha$-dependent *FSE*, i.e., Eqs. (8) and (9), in order to determine $\mathbf{h}_u$ for all possible vectors $\mathbf{h}_\alpha$.

In most practical situations, the number of model parameters exceeds significantly the number of functional responses of interest, i.e., $N_r \ll N_\alpha$. In such cases, *the ASAP is the most efficient method for computing exactly the first-order sensitivities since it requires only $O(N_r)$ large-scale computations*. To implement the *ASAP* for computing the first-order G-differential $\delta R(\mathbf{e}^0;\mathbf{h})$, the spaces $\mathscr{E}_u$ and $\mathscr{E}_Q$ will henceforth be considered to be Hilbert spaces and denoted as $\mathscr{H}_u(\Omega_x)$ and $\mathscr{H}_Q(\Omega_x)$, respectively. The elements of $\mathscr{H}_u(\Omega_x)$ and $\mathscr{H}_Q(\Omega_x)$ are, as before, vector-valued functions defined on the open set $\Omega_x \subset \mathbb{R}^{J_x}$, with smooth boundary $\partial \Omega_x$. On $\mathscr{H}_u(\Omega_x)$, the inner product of two vectors $\mathbf{u}^{(1)} \in \mathscr{H}_u$ and $\mathbf{u}^{(2)} \in \mathscr{H}_u$ will be denoted as $\langle \mathbf{u}^{(1)}, \mathbf{u}^{(2)} \rangle_u$, while the inner product [on $\mathscr{H}_Q(\Omega_x)$] of two vectors $\mathbf{Q}^{(1)} \in \mathscr{H}_Q$ and $\mathbf{Q}^{(2)} \in \mathscr{H}_Q$ will be denoted as $\langle \mathbf{Q}^{(1)}, \mathbf{Q}^{(2)} \rangle_Q$. Furthermore, the *ASAP* also requires that $\delta R(\mathbf{e}^0;\mathbf{h})$ be linear in $\mathbf{h}$, which implies that $R(\mathbf{e})$ must satisfy a weak Lipschitz condition at $\mathbf{e}^0$, and also satisfy the following condition

$$R(\mathbf{e}^0 + \varepsilon \mathbf{h}_1 + \varepsilon \mathbf{h}_2) - R(\mathbf{e}^0 + \varepsilon \mathbf{h}_1) - R(\mathbf{e}^0 + \varepsilon \mathbf{h}_2) + R(\mathbf{e}^0) = o(t);$$
$$\mathbf{h}_1, \mathbf{h}_2 \in \mathscr{H}_u \times \mathscr{H}_\alpha; \quad \varepsilon \in \mathscr{F}. \tag{10}$$

If $R(\mathbf{e})$ satisfies the two conditions above, then the total response variation $\delta R(\mathbf{e}^0;\mathbf{h})$ is indeed linear in $\mathbf{h}$, and can therefore be denoted as $DR(\mathbf{e}^0;\mathbf{h})$. Consequently, $R(\mathbf{e})$ admits a total G-derivative at $\mathbf{e}^0 = (\mathbf{u}^0, \boldsymbol{\alpha}^0)$, such that the relationship

$$DR(\mathbf{e}^0;\mathbf{h}) = R'_u(\mathbf{e}^0)\mathbf{h}_u + R'_\alpha(\mathbf{e}^0)\mathbf{h}_\alpha \tag{11}$$

holds, where $R'_u(\mathbf{e}^0)$ and $R'_\alpha(\mathbf{e}^0)$ are the partial G-derivatives at $\mathbf{e}^0$ of $R(\mathbf{e})$ with respect to $\mathbf{u}$ and $\boldsymbol{\alpha}$. It is convenient to refer to the quantities $R'_u(\mathbf{e}^0)\mathbf{h}_u$ and $R'_\alpha(\mathbf{e}^0)\mathbf{h}_\alpha$ appearing in Eq. (11) as the "*indirect effect term*" and the "*direct effect term*," respectively. The operator $R'_u(\mathbf{e}^0)$ acts linearly on the vector of (arbitrary) variations $\mathbf{h}_u$, from $\mathcal{H}_u$ into $\mathcal{F}$, while the operator $R'_\alpha(\mathbf{e}^0)$ acts linearly on the vector of (arbitrary) variations $\mathbf{h}_\alpha$, from $\mathcal{H}_u$ into $\mathcal{F}$. Since the functional $R'_u(\mathbf{e}^0)\mathbf{h}_u$ is linear in $\mathbf{h}_u$ and since Hilbert spaces are self-dual, the Riesz representation theorem ensures that there exists a unique column vector $\mathbf{D}_u R(\mathbf{e}^0) \in \mathcal{H}_u$, where $\mathbf{D}_u R(\mathbf{e}) \equiv [\partial R(\mathbf{e})/\partial u_1, \ldots, \partial R(\mathbf{e})/\partial u_{K_u}]$, which is customarily called the partial gradient of $R(\mathbf{e})$ with respect to $\mathbf{u}$, evaluated at $\mathbf{e}^0$, such that

$$R'_u(e^0)\mathbf{h}_u = \langle \mathbf{D}_u R(e^0), \mathbf{h}_u \rangle_u, \quad \mathbf{h}_u \in \mathcal{H}_u. \tag{12}$$

Similarly, the functional $R'_\alpha(\mathbf{e}^0)\mathbf{h}_\alpha$ is linear in $\mathbf{h}_\alpha$ and since Hilbert spaces are self-dual, the Riesz representation theorem ensures that there exists a unique vector $\mathbf{D}_\alpha R(e^0) \in \mathcal{H}_\alpha$, where $\mathbf{D}_\alpha R(e) \equiv [\partial R(e)/\partial \alpha_1, \ldots, \partial R(e)/\partial \alpha_{N_\alpha}]$, which is customarily called the partial gradient of $R(\mathbf{e})$ with respect to $\boldsymbol{\alpha}$, evaluated at $\mathbf{e}^0$, such that

$$R'_\alpha(e^0)\mathbf{h}_\alpha = \langle \mathbf{D}_\alpha R(e^0), \mathbf{h}_\alpha \rangle_\alpha, \quad \mathbf{h}_\alpha \in \mathcal{H}_\alpha, \tag{13}$$

where the inner product in the Hilbert space $\mathcal{H}_\alpha$ is denoted as $\langle \bullet, \bullet \rangle_\alpha$.

Following [1], we construct the *ASAP* by introducing formal adjoint $\mathbf{L}^*(\boldsymbol{\alpha}^0)$ of $\mathbf{L}(\boldsymbol{\alpha}^0)$ and recalling from the geometry of Hilbert spaces that the following relationship holds for an arbitrary vector $\boldsymbol{\psi} \in \mathcal{H}_Q$:

$$\left\langle \boldsymbol{\psi}, \mathbf{L}(\boldsymbol{\alpha}^0)\mathbf{h}_u \right\rangle_Q = \left\langle \mathbf{L}^*(\boldsymbol{\alpha}^0)\boldsymbol{\psi}, \mathbf{h}_u \right\rangle_u + \left\{P(\mathbf{h}_u;\boldsymbol{\psi})\right\}_{\partial\Omega_x} \quad (14)$$

In the above equation, the formal adjoint operator $\mathbf{L}^*(\boldsymbol{\alpha}^0)$ is the $K_u \times K_u$ matrix

$$\mathbf{L}^*(\boldsymbol{\alpha}^0) \equiv \left[L_{ji}^*(\boldsymbol{\alpha}^0)\right], \ (i, j = 1, \ldots, K_u), \quad (15)$$

comprising elements $L_{ji}^*(\boldsymbol{\alpha}^0)$ obtained by transposing the formal adjoints of the operators $L_{ij}(\boldsymbol{\alpha}^0)$, while $\left\{P(\mathbf{h}_u;\boldsymbol{\psi})\right\}_{\partial\Omega_x}$ is the associated bilinear form evaluated on $\partial\Omega_x$. The domain of $\mathbf{L}^*(\mathbf{e}^0)$ is determined by selecting appropriate adjoint boundary and/or initial conditions, represented here in operator form as

$$\mathbf{B}^*(\boldsymbol{\alpha}^0)\boldsymbol{\psi}^0 - \mathbf{A}^*(\boldsymbol{\alpha}^0) = \mathbf{0}, \ \mathbf{x} \in \partial\Omega_x. \quad (16)$$

The above boundary conditions for $\mathbf{L}^*(\mathbf{e}^0)$ are obtained by requiring that:
  (a) They must be independent of *unknown* values of $\mathbf{h}_u$ and $\mathbf{h}_\alpha$;
  (b) The substitution of the boundary and/or initial conditions represented by Eqs. (9) and (16) into the expression of $\left\{P(\mathbf{h}_u;\boldsymbol{\psi})\right\}_{\partial\Omega_x}$ must cause all terms containing unknown values of $\mathbf{h}_u$ to vanish.

This selection of the boundary conditions for $\mathbf{L}^*(\mathbf{e}^0)$ reduces the bilinear concomitant $\left\{P(\mathbf{h}_u;\boldsymbol{\psi})\right\}_{\partial\Omega_x}$ to a quantity that contains boundary terms involving only known values of $\mathbf{h}_\alpha$, $\boldsymbol{\psi}$, and, possibly, $\boldsymbol{\alpha}^0$; this quantity will be denoted by $\hat{P}(\mathbf{h}_\alpha,\boldsymbol{\psi};\boldsymbol{\alpha}^0)$. In general, $\hat{P}$ does not automatically vanish as a result of these manipulations, although it may do so in particular instances. In principle, $\hat{P}$ could be forced to vanish by considering extensions of $\mathbf{L}(\boldsymbol{\alpha}^0)$, in the operator sense, but this is seldom needed in practice.

Introducing now Eqs. (8), (9), and (16) into Eq. (14) and re-arranging the resulting equation yields

$$\left\langle \mathbf{L}^*(\boldsymbol{\alpha}^0)\boldsymbol{\psi}, \mathbf{h}_u \right\rangle_u = \left\langle \boldsymbol{\psi}, \left\langle \mathbf{D}_\alpha \mathbf{Q}(\boldsymbol{\alpha}^0) - \mathbf{D}_\alpha \mathbf{L}(\boldsymbol{\alpha}^0)\mathbf{u}^0, \mathbf{h}_\alpha \right\rangle_\alpha \right\rangle_Q - \hat{P}(\mathbf{h}_\alpha, \boldsymbol{\psi}; \boldsymbol{\alpha}^0). \tag{17}$$

Since $\boldsymbol{\psi}$ is not completely defined yet, we now complete its definition by requiring that the left-side of Eq. (17) and the right-side of Eq. (12) represent the same functional, which is accomplished by imposing the relationship

$$\mathbf{L}^*(\boldsymbol{\alpha}^0)\boldsymbol{\psi}^0 = \mathbf{D}_u R(\mathbf{e}^0), \tag{18}$$

where the superscript "zero" emphasizes the fact that the function $\boldsymbol{\psi}^0$ satisfies Eqs. (18) and (16) for the nominal parameter values $\boldsymbol{\alpha}^0$. Note that the well-known Riesz representation theorem ensures that the above relationship, where $\boldsymbol{\psi}^0$ satisfies the adjoint boundary conditions given in Eq. (16), holds uniquely. The construction of the requisite adjoint system, consisting of Eqs. (18) and (16), has thus been accomplished. Furthermore, Eqs. (11) through (18) can now be used to obtain the following expression for the total sensitivity $DR(\mathbf{e}^0; \mathbf{h})$ of $R(\mathbf{e})$ at:

$$DR(\mathbf{e}^0; \mathbf{h}) = \left\langle \mathbf{D}_\alpha R(\mathbf{e}^0), \mathbf{h}_\alpha \right\rangle_\alpha + \left\langle \boldsymbol{\psi}^0, \left\langle \mathbf{D}_\alpha \mathbf{Q}(\boldsymbol{\alpha}^0) - \mathbf{D}_\alpha \mathbf{L}(\boldsymbol{\alpha}^0)\mathbf{u}^0, \mathbf{h}_\alpha \right\rangle_\alpha \right\rangle_\psi - \hat{P}(\mathbf{h}_\alpha, \boldsymbol{\psi}^0; \boldsymbol{\alpha}^0)$$
$$\equiv \mathbf{S}(\mathbf{u}^0, \boldsymbol{\alpha}^0, \boldsymbol{\psi}^0)\mathbf{h}_\alpha = \sum_{i=1}^{N_\alpha} S_i(\mathbf{u}^0, \boldsymbol{\alpha}^0, \boldsymbol{\psi}^0)\delta\alpha_i,$$
$$\tag{19}$$

where $\mathbf{S}(\mathbf{u}^0, \boldsymbol{\alpha}^0, \boldsymbol{\psi}^0) \equiv (S_1, ..., S_{N_\alpha})^\dagger$, and where the $i^{th}$-partial first-order sensitivity (G-derivative), $S_i(\mathbf{u}^0, \boldsymbol{\alpha}^0, \boldsymbol{\psi}^0)$, of $R(\mathbf{e})$ with respect to the $i^{th}$-model parameter $\alpha_i$, $i = 1, ..., N_\alpha$, is given by the expression

$$S_i\left(\mathbf{u}^0,\boldsymbol{\alpha}^0,\boldsymbol{\psi}^0\right) \equiv \left(\mathbf{D}_\alpha R\right)_i + \left\langle \boldsymbol{\psi}^0, \left[\mathbf{D}_\alpha \mathbf{Q}\left(\boldsymbol{\alpha}^0\right) - \mathbf{D}_\alpha \mathbf{L}\left(\boldsymbol{\alpha}^0\right)\mathbf{u}^0\right]_i \right\rangle_\psi - \left[\mathbf{D}_\alpha \hat{P}\left(\mathbf{u}^0,\boldsymbol{\psi}^0,\boldsymbol{\alpha}^0\right)\right]_i$$

$$= \frac{\partial R(\mathbf{u},\boldsymbol{\alpha})}{\partial \alpha_i}\bigg|_{(\mathbf{u}^0,\boldsymbol{\alpha}^0)} + \left\{\left\langle \boldsymbol{\psi}, \frac{\partial \mathbf{Q}(\mathbf{u},\boldsymbol{\alpha})}{\partial \alpha_i} - \frac{\partial[\mathbf{L}(\boldsymbol{\alpha})\mathbf{u}]}{\partial \alpha_i} \right\rangle_\psi\right\}_{(\mathbf{u}^0,\boldsymbol{\psi}^0;\boldsymbol{\alpha}^0)} - \frac{\partial \hat{P}(\mathbf{u},\boldsymbol{\psi},\boldsymbol{\alpha})}{\partial \alpha_i}\bigg|_{(\mathbf{u}^0,\boldsymbol{\psi}^0;\boldsymbol{\alpha}^0)}, \quad i=1,\ldots,N_\alpha.$$

(20)

All partial derivatives in the above expressions are to be understood as partial G-derivatives, of course. As Eq. (20) indicates, the desired elimination of all unknown values of $\mathbf{h}_u$ from the expressions of the sensitivities $S_i(\mathbf{u},\boldsymbol{\alpha},\boldsymbol{\psi})$, $i=1,\ldots,N_\alpha$, of $R(\mathbf{e})$ at $\mathbf{e}^0$ has been accomplished. Note that we have designated the space $\mathcal{H}_Q(\Omega_x)$ as $\mathcal{H}_\psi(\Omega_x)$, in order to emphasize that we are dealing with the Hilbert space on which the adjoint function $\boldsymbol{\psi}$ is defined. *The sensitivities $S_i(\mathbf{u},\boldsymbol{\alpha},\boldsymbol{\psi})$ can therefore be computed by means of Eq. (20), after solving **only once** the adjoint sensitivity system, consisting of Eqs. (16) and (18), to obtain the adjoint function $\boldsymbol{\psi}$.* It is very important to note that this adjoint system is independent not only of the functions $\mathbf{h}_u$ but also of the nominal values $\mathbf{u}^0$ of $\mathbf{u}$. This means that the adjoint system, namely Eqs. (16) and (18), can be solved independently of the solution $\mathbf{u}^0$ of the original equations. In turn, this fact simplifies considerably the choice of numerical methods for solving the adjoint system. *It is also important to note that this advantageous situation arises if and only if the original equations that model the physical problem, i.e., Eqs. (1) and (2), are linear in the state-variable $\mathbf{u}$.*

## 3. THE SECOND-ORDER FORWARD AND ADJOINT SENSITIVITY ANALYSIS PROCEDURES FOR LARGE-SCALE LINEAR SYSTEMS

We now note the very important fact that, since Eq. (20) holds for any nominal parameter values, $\boldsymbol{\alpha}^0$, it follows that the first-order sensitivities $S_i(\mathbf{u},\boldsymbol{\alpha},\boldsymbol{\psi})$, $i=1,\ldots,N_\alpha$, can be generally considered as functionals of the original state-function $\mathbf{u}$, the parameters $\boldsymbol{\alpha}$, and the adjoint function $\boldsymbol{\psi}$, namely

$$S_i(\mathbf{u},\boldsymbol{\alpha},\boldsymbol{\psi}) = \frac{\partial R(\mathbf{u},\boldsymbol{\alpha})}{\partial \alpha_i} + \left\langle \boldsymbol{\psi}, \frac{\partial Q(\mathbf{u},\boldsymbol{\alpha})}{\partial \alpha_i} - \frac{\partial[\mathbf{L}(\boldsymbol{\alpha})\mathbf{u}]}{\partial \alpha_i} \right\rangle_\psi - \frac{\partial \hat{P}(\mathbf{u},\boldsymbol{\psi},\boldsymbol{\alpha})}{\partial \alpha_i}, \quad i=1,\ldots,N_\alpha, \quad (21)$$

where the original state-function **u** satisfies Eqs. (1) and (2), while the adjoint function **ψ** satisfies Eqs. (16) and (18) for any physically-defined values of the parameters **α**, i.e.,

$$\mathbf{L}^*(\boldsymbol{\alpha})\boldsymbol{\psi} = \mathbf{D}_u R(\mathbf{u},\boldsymbol{\alpha}), \quad \mathbf{x}\in\Omega_x, \quad (22)$$

together with the corresponding adjoint boundary and/or initial conditions

$$\mathbf{B}^*(\boldsymbol{\alpha})\boldsymbol{\psi} - \mathbf{A}^*(\boldsymbol{\alpha}) = \mathbf{0}, \quad \mathbf{x}\in\partial\Omega_x. \quad (23)$$

## 3.1. The Second-Order Forward Sensitivity Analysis Procedure (SO-FSAP) for Large-Scale Linear Systems

Hence, as Eq. (21) indicates, the first-order sensitivities are functionals of the form $S_i(\mathbf{u},\boldsymbol{\alpha},\boldsymbol{\psi}):\mathcal{D}_{R_i}\subset\mathcal{H}_u(\Omega_x)\times\mathcal{H}_\alpha\times\mathcal{H}_\psi(\Omega_x)\to\mathcal{F}$. Hence, it is possible to define the first-order G-differential, $\delta S_i(\mathbf{e}^0,\boldsymbol{\psi}^0,\mathbf{g})$, of any of the functionals $S_i(\mathbf{u},\boldsymbol{\alpha},\boldsymbol{\psi})$, at the point $(\mathbf{e}^0,\boldsymbol{\psi}^0)$ in the usual manner, namely

$$\delta S_i(\mathbf{e}^0,\boldsymbol{\psi}^0;\mathbf{g}) \equiv \left\{\frac{d}{d\varepsilon}\left[S_i(\mathbf{u}^0+\varepsilon\mathbf{h}_u,\boldsymbol{\psi}^0+\varepsilon\mathbf{h}_\psi,\boldsymbol{\alpha}^0+\varepsilon\mathbf{h}_\alpha)\right]\right\}_{\varepsilon=0}, \quad (24)$$

For an arbitrary scalar $\varepsilon\in\mathcal{F}$, and all (i.e., arbitrary) vectors $\mathbf{g}\equiv(\mathbf{h}_u,\mathbf{h}_\alpha,\mathbf{h}_\psi)\in\mathcal{H}_u(\Omega_x)\times\mathcal{H}_\alpha\times\mathcal{H}_\psi(\Omega_x)$. Applying the above definition to the expression of $S_i(\mathbf{u},\boldsymbol{\alpha},\boldsymbol{\psi})$ given by Eq. (21) yields

$$\delta S_i(\mathbf{e}^0,\boldsymbol{\psi}^0;\mathbf{g}) = (\delta S_i)_{direct} + (\delta S_i)_{indirect}, \quad (25)$$

where $(\delta S_i)_{direct}$ denotes the "*direct-effect term*" and $(\delta S_i)_{indirect}$ denotes the "*indirect-effect term.*" The "*direct-effect term,*" $(\delta S_i)_{direct}$, is defined as

$$
\begin{aligned}
(\delta S_i)_{direct} &\equiv \langle \mathbf{D}_\alpha S_i, \mathbf{h}_\alpha \rangle_\alpha \\
&= \frac{\partial}{\partial \alpha_i} \left\{ \left\langle \mathbf{D}_\alpha R(\mathbf{u},\boldsymbol{\alpha}) - \mathbf{D}_\alpha \hat{P}(\mathbf{u},\boldsymbol{\psi},\boldsymbol{\alpha}) + \left\langle \boldsymbol{\psi}, \mathbf{D}_\alpha \mathbf{Q}(\boldsymbol{\alpha}) - \mathbf{D}_\alpha [\mathbf{L}(\boldsymbol{\alpha})\mathbf{u}] \right\rangle_\psi, \mathbf{h}_\alpha \right\rangle_\alpha \right\}_{(\mathbf{u}^0, \boldsymbol{\psi}^0; \boldsymbol{\alpha}^0)},
\end{aligned}
\tag{26}
$$

and *can be computed immediately* at this stage, without needing any additional large-scale computations. On the other hand, the "*indirect-effect term*", $(\delta S_i)_{indirect}$, is defined as

$$
(\delta S_i)_{indirect} \equiv \left\{ (\mathbf{h}_u, \mathbf{h}_\psi)^\dagger \begin{pmatrix} \mathbf{D}_u S_i(\mathbf{u},\boldsymbol{\psi},\boldsymbol{\alpha}) \\ \mathbf{D}_\psi S_i(\mathbf{u},\boldsymbol{\psi},\boldsymbol{\alpha}) \end{pmatrix} \right\}_{(\mathbf{u}^0, \boldsymbol{\psi}^0; \boldsymbol{\alpha}^0)},
\tag{27}
$$

where

$$
\mathbf{D}_\psi S_i(\mathbf{u},\boldsymbol{\psi},\boldsymbol{\alpha}) \equiv \frac{\partial}{\partial \alpha_i} \left[ \mathbf{Q}(\boldsymbol{\alpha}) - \mathbf{L}(\boldsymbol{\alpha})\mathbf{u} - \mathbf{D}_\psi \hat{P}(\mathbf{u},\boldsymbol{\psi},\boldsymbol{\alpha}) \right],
\tag{28}
$$

$$
\mathbf{D}_u S_i(\mathbf{u},\boldsymbol{\psi},\boldsymbol{\alpha}) \equiv \frac{\partial}{\partial \alpha_i} \left[ \mathbf{D}_u R(\mathbf{u},\boldsymbol{\alpha}) - \mathbf{D}_u \hat{P}(\mathbf{u},\boldsymbol{\psi},\boldsymbol{\alpha}) - \Lambda(\boldsymbol{\psi},\boldsymbol{\alpha}) \right],
\tag{29}
$$

with

$$
\Lambda(\boldsymbol{\psi},\boldsymbol{\alpha}) \equiv (\Lambda_1, \ldots, \Lambda_{K_u}); \quad \Lambda_k(\boldsymbol{\psi},\boldsymbol{\alpha}) \equiv \sum_{j=1}^{K_u} \psi_j(\mathbf{x}) \frac{\partial L_{jk}(\boldsymbol{\alpha})}{\partial \alpha_k}; \quad k = 1, \ldots, K_u.
\tag{30}
$$

Note that the "indirect-effect term" *cannot be computed at this stage*, since the vectors of variations $\mathbf{h}_u$ and $\mathbf{h}_\psi$ are unknown. Recall that the vector $\mathbf{h}_u$ is the $\mathbf{h}_\alpha$–dependent solution of the *forward sensitivity system* expressed by Eqs. (8) and (9), which represent large-scale systems that are computationally impractical to solve for large-scale systems with many parameters. On the other hand, the vector of variations $\mathbf{h}_\psi$ around the nominal value $\boldsymbol{\psi}^0$ will be the solution of the system of operator equations that will result from applying the definition of the G-differential to the adjoint sensitivity system Eqs. (22) and (23), namely

$$\left[\frac{d}{d\varepsilon}\left\{\mathbf{L}^{*}\left[\left(\boldsymbol{\alpha}^{0}+\varepsilon\mathbf{h}_{\alpha}\right)\right]\left(\boldsymbol{\psi}^{0}+\varepsilon\mathbf{h}_{\psi}\right)-\mathbf{D}_{u}R\left(\mathbf{u}^{0}+\varepsilon\mathbf{h}_{u},\boldsymbol{\alpha}^{0}+\varepsilon\mathbf{h}_{\alpha}\right)\right\}\right]_{\varepsilon=0}=\mathbf{0},\ \mathbf{x}\in\Omega_{x}, \quad (31)$$

$$\left[\frac{d}{d\varepsilon}\left\{\mathbf{B}^{*}\left[\left(\boldsymbol{\alpha}^{0}+\varepsilon\mathbf{h}_{\alpha}\right)\right]\left(\boldsymbol{\psi}^{0}+\varepsilon\mathbf{h}_{\psi}\right)-\mathbf{A}^{*}\left(\boldsymbol{\alpha}^{0}+\varepsilon\mathbf{h}_{\alpha}\right)\right\}\right]_{\varepsilon=0}=\mathbf{0},\ \mathbf{x}\in\partial\Omega_{x}. \quad (32)$$

As the above equations clearly show, the vector of variations $\mathbf{h}_{\psi}$ (around the nominal value $\boldsymbol{\psi}^{0}$) will depend on the parameters variations $\mathbf{h}_{\alpha}$. Together, the forward sensitivity equations, namely Eqs. (8) and (9), and the G-differentiated adjoint sensitivity system, namely Eqs. (32) and (33), can be written in the following block-matrix-operator form:

$$\begin{pmatrix} \mathbf{L}(\boldsymbol{\alpha}^{0}) & \mathbf{0} \\ -\mathbf{D}_{uu}^{2}R(\boldsymbol{e}^{0}) & \mathbf{L}^{*}(\boldsymbol{\alpha}^{0}) \end{pmatrix}\begin{pmatrix} \mathbf{h}_{u} \\ \mathbf{h}_{\psi} \end{pmatrix}=\begin{pmatrix} \left\langle \mathbf{D}_{\alpha}\mathbf{Q}(\boldsymbol{\alpha}^{0})-\mathbf{D}_{\alpha}\left[\mathbf{L}(\boldsymbol{\alpha}^{0})\mathbf{u}^{0}\right],\mathbf{h}_{\alpha}\right\rangle_{\alpha} \\ \left\langle \mathbf{D}_{\alpha u}^{2}R(\boldsymbol{e}^{0})-\mathbf{L}^{*}\left[(\boldsymbol{\alpha}^{0})\boldsymbol{\psi}^{0}\right],\mathbf{h}_{\alpha}\right\rangle_{\alpha} \end{pmatrix},\ \mathbf{x}\in\Omega_{x}, \quad (33)$$

together with the corresponding G-differentiated boundary and/or initial conditions

$$\begin{pmatrix} \mathbf{B}(\boldsymbol{\alpha}^{0}) & \mathbf{0} \\ \mathbf{0} & \mathbf{B}^{*}(\boldsymbol{\alpha}^{0}) \end{pmatrix}\begin{pmatrix} \mathbf{h}_{u} \\ \mathbf{h}_{\psi} \end{pmatrix}=\begin{pmatrix} \left\langle \mathbf{D}_{\alpha}\mathbf{A}(\boldsymbol{\alpha}^{0})-\mathbf{D}_{\alpha}\left[\mathbf{B}(\boldsymbol{\alpha}^{0})\mathbf{u}^{0}\right],\mathbf{h}_{\alpha}\right\rangle_{\alpha} \\ \left\langle \mathbf{D}_{\alpha}\mathbf{A}^{*}(\boldsymbol{\alpha}^{0})-\mathbf{D}_{\alpha}\left[\mathbf{B}^{*}(\boldsymbol{\alpha}^{0})\boldsymbol{\psi}^{0}\right],\mathbf{h}_{\alpha}\right\rangle_{\alpha} \end{pmatrix},\ \mathbf{x}\in\partial\Omega_{x}. \quad (34)$$

In Eq. (33), the operators $\mathbf{D}_{uu}^{2}R(\boldsymbol{e}^{0})$ and, respectively, $\mathbf{D}_{\alpha u}^{2}R(\boldsymbol{e}^{0})$ are matrices of partial G-derivatives, of the form

$$\mathbf{D}_{\alpha u}^{2}R(\boldsymbol{e})\equiv\begin{pmatrix} \frac{\partial^{2}R(\boldsymbol{e})}{\partial u_{1}\partial\alpha_{1}} & \cdots & \frac{\partial^{2}R(\boldsymbol{e})}{\partial u_{1}\partial\alpha_{N_{\alpha}}} \\ \vdots & \ddots & \vdots \\ \frac{\partial^{2}R(\boldsymbol{e})}{\partial u_{N_{u}}\partial\alpha_{1}} & \cdots & \frac{\partial^{2}R(\boldsymbol{e})}{\partial u_{N_{u}}\partial\alpha_{N_{\alpha}}} \end{pmatrix};\ \mathbf{D}_{uu}^{2}R(\boldsymbol{e})\equiv\begin{pmatrix} \frac{\partial^{2}R(\boldsymbol{e})}{\partial u_{1}^{2}} & \cdots & \frac{\partial^{2}R(\boldsymbol{e})}{\partial u_{1}\partial u_{N_{u}}} \\ \vdots & \ddots & \vdots \\ \frac{\partial^{2}R(\boldsymbol{e})}{\partial u_{N_{u}}\partial u_{1}} & \cdots & \frac{\partial^{2}R(\boldsymbol{e})}{\partial u_{N_{u}}^{2}} \end{pmatrix}. \quad (35)$$

The block matrix Eq. (33) together with the boundary and/or initial conditions represented by Eq. (34) constitute the *second-order forward sensitivity system*. These equations can be solved to obtain the vectors $\mathbf{h}_{u}$ and $\mathbf{h}_{\psi}$, which can, in turn, be used in Eq. (27) to compute the indirect-effect term, $(\delta S_{i})_{indirect}$. In view of Eq. (25), this indirect effect term would then be added together with the already computed direct-effect term, $(\delta S_i)_{direct}$, to obtain the vector, $S_i(\mathbf{u},\boldsymbol{\alpha},\boldsymbol{\psi})$, of partial mixed second-order sensitivities of the form $\{\partial^2 R(\mathbf{u},\boldsymbol{\alpha})/\partial \alpha_j \partial \alpha_i\}_{(\mathbf{u}^0,\boldsymbol{\alpha}^0)}$, $j=1,...,N_\alpha$. Thus, starting to solve Eqs. (33) and (34) for $S_1(\mathbf{u},\boldsymbol{\alpha},\boldsymbol{\psi})$, *for* $i=1$, the process of solving these operator equations for $S_2(\mathbf{u},\boldsymbol{\alpha},\boldsymbol{\psi})$, *for* $i=2$, would be continued/repeated until ending with $S_{N_\alpha}(\mathbf{u},\boldsymbol{\alpha},\boldsymbol{\psi})$, *for* $i=N_\alpha$.

Clearly, the computational process just described would yield the complete set of all second-order responses sensitivities, $\{\partial^2 R(\mathbf{u},\boldsymbol{\alpha})/\partial \alpha_j \partial \alpha_i\}_{(\mathbf{u}^0,\boldsymbol{\alpha}^0)}$, to all of the system parameters, parameters, in $O(N_\alpha^2)$ large-scale forward computations. Of course, such a computational burden would be impractical for large-scale systems, with millions of parameters, as are routinely encountered in neutron/photon transport problems. Note that although the forward sensitivity system and the G-differentiated adjoint system have been written in block matrix form [cf., Eqs. (33) and (34)] as though they were coupled, they are actually not coupled and can be solved independently of one another. *This uncoupling occurs only when the original system is linear in the state-function* $\mathbf{u}$; *for nonlinear systems in* $\mathbf{u}$, *Eqs. (33) and (34) would be coupled, in general.*

### *3.2. The Second-Order Adjoint Sensitivity Analysis Procedure (SO-ASAP) for Large-Scale Linear Systems*

The large number of large-scale computations involved in the *SO-FSAP* are needed for the computation of the indirect-effect term, $(\delta S_i)_{indirect}$, just as was the case in Section 2, where the first-order ASAP was reviewed. The second-order adjoint sensitivity analysis procedure (*SO-ASAP)* will aim precisely at avoiding the forward computation of $(\delta S_i)_{indirect}$, by computing it, alternatively, using an appropriately developed *adjoint system.* For this purpose, consider now two vectors $\mathbf{v}^{(1)} \equiv (\mathbf{u}^{(1)}, \boldsymbol{\psi}^{(1)}) \in \mathcal{H}_{u\psi}$ and $\mathbf{v}^{(2)} \equiv (\mathbf{u}^{(2)}, \boldsymbol{\psi}^{(2)}) \in \mathcal{H}_{u\psi}$ in the Hilbert space $\mathcal{H}_{u\psi}(\Omega_x) \equiv \mathcal{H}_u(\Omega_x) \times \mathcal{H}_\psi(\Omega_x)$, endowed with the inner product $\langle \mathbf{v}^{(1)}, \mathbf{v}^{(2)} \rangle_{u\psi}$ defined as

$$\langle \mathbf{v}^{(1)}, \mathbf{v}^{(2)} \rangle_{u\psi} \equiv \int_{\Omega_x} \left( \mathbf{u}^{(1)} \cdot \mathbf{u}^{(2)} + \mathbf{\psi}^{(1)} \cdot \mathbf{\psi}^{(2)} \right) d\mathbf{x}. \tag{36}$$

Next, we form the above inner product of Eq. (33) with a yet undefined vector $\left( \mathbf{\psi}^{(1)}, \mathbf{\psi}^{(2)} \right)^\dagger$ to obtain the sequence of equalities shown below:

$$\left\langle \left( \mathbf{\psi}^{(1)}, \mathbf{\psi}^{(2)} \right) \begin{pmatrix} \mathbf{L}(\boldsymbol{\alpha}^0) & 0 \\ -\mathbf{D}_{uu}^2 R(e^0) & \mathbf{L}^*(\boldsymbol{\alpha}^0) \end{pmatrix} \begin{pmatrix} \mathbf{h}_u \\ \mathbf{h}_\psi \end{pmatrix} \right\rangle_{u\psi}$$

$$= \left\langle \left( \mathbf{\psi}^{(1)}, \mathbf{\psi}^{(2)} \right) \begin{pmatrix} \left\langle \mathbf{D}_\alpha \mathbf{Q}(\boldsymbol{\alpha}^0) - \mathbf{D}_\alpha \left[ \mathbf{L}(\boldsymbol{\alpha}^0) \mathbf{u}^0 \right], \mathbf{h}_\alpha \right\rangle_\alpha \\ \left\langle \mathbf{D}_{\alpha u}^2 R(e^0) - \mathbf{L}^* \left[ (\boldsymbol{\alpha}^0) \mathbf{\psi}^0 \right], \mathbf{h}_\alpha \right\rangle_\alpha \end{pmatrix} \right\rangle_{u\psi}$$

$$= \left\langle (\mathbf{h}_u, \mathbf{h}_\psi) \begin{pmatrix} \mathbf{L}^*(\boldsymbol{\alpha}^0) & -\left[ \mathbf{D}_{uu}^2 R(e^0) \right]^* \\ 0 & \mathbf{L}(\boldsymbol{\alpha}^0) \end{pmatrix} \begin{pmatrix} \mathbf{\psi}^{(1)} \\ \mathbf{\psi}^{(2)} \end{pmatrix} \right\rangle_{u\psi} + \left\{ P_2 \left( \mathbf{u}^0, \boldsymbol{\alpha}^0; \mathbf{h}_u, \mathbf{h}_\psi, \mathbf{h}_\alpha; \mathbf{\psi}^{(1)}, \mathbf{\psi}^{(2)} \right) \right\}_{\partial \Omega_x}, \quad \mathbf{x} \in \Omega_x,$$

$$\tag{37}$$

where $\left[ \mathbf{D}_{uu}^2 R(e^0) \right]^*$ denotes the formal adjoint of $\mathbf{D}_{uu}^2 R(e^0)$ and $\left\{ P_2 \left( \mathbf{u}^0, \boldsymbol{\alpha}^0; \mathbf{h}_u, \mathbf{h}_\psi, \mathbf{h}_\alpha; \mathbf{\psi}^{(1)}, \mathbf{\psi}^{(2)} \right) \right\}_{\partial \Omega_x}$ denotes corresponsing the bilinear concomitant on $\mathbf{x} \in \partial \Omega_x$. Folowing the same principles as in Section 2, we require the first term on the right-side of the last equality in Eq. (37) to represent the same functional as the right-side of Eq. (27), which yields the block-matrix equation

$$\begin{pmatrix} \mathbf{L}^*(\boldsymbol{\alpha}^0) & -\left[ \mathbf{D}_{uu}^2 R(e^0) \right]^* \\ 0 & \mathbf{L}(\boldsymbol{\alpha}^0) \end{pmatrix} \begin{pmatrix} \mathbf{\psi}^{(1)} \\ \mathbf{\psi}^{(2)} \end{pmatrix} = \begin{pmatrix} \mathbf{D}_u S_i(\mathbf{u}^0, \mathbf{\psi}^0, \boldsymbol{\alpha}^0) \\ \mathbf{D}_\psi S_i(\mathbf{u}^0, \mathbf{\psi}^0, \boldsymbol{\alpha}^0) \end{pmatrix} \tag{38.a}$$

or, in component form

$$\mathbf{L}(\boldsymbol{\alpha}^0)\boldsymbol{\psi}^{(2)} = \mathbf{D}_\psi S_i(\mathbf{u}^0, \boldsymbol{\psi}^0, \boldsymbol{\alpha}^0),$$
$$\mathbf{L}^*(\boldsymbol{\alpha}^0)\boldsymbol{\psi}^{(1)} = \mathbf{D}_u S_i(\mathbf{u}^0, \boldsymbol{\psi}^0, \boldsymbol{\alpha}^0) + \left[\mathbf{D}_{uu}^2 R(e^0)\right]^* \boldsymbol{\psi}^{(2)}.$$
(38.b)

The definition of the functions $\boldsymbol{\psi}^{(1)}$ and $\boldsymbol{\psi}^{(2)}$ can now be completed by requiring them to satisfy adjoint boundary conditions denoted as

$$\mathbf{B}_1^*(\boldsymbol{\psi}^{(1)}, \boldsymbol{\psi}^{(2)}, \boldsymbol{\alpha}^0) = \mathbf{0}, \quad \mathbf{x} \in \partial\Omega_x,$$
$$\mathbf{B}_2^*(\boldsymbol{\psi}^{(1)}, \boldsymbol{\psi}^{(2)}, \boldsymbol{\alpha}^0) = \mathbf{0}, \quad \mathbf{x} \in \partial\Omega_x.$$
(39)

The above adjoint boundary and/or initial are determined following the same principles as described in Section 2, namely that:

(a) They must be independent of *unknown* values of $\mathbf{h}_u, \mathbf{h}_\psi$, and $\mathbf{h}_\alpha$;

(b) The substitution of the boundary and/or initial conditions represented by Eqs. (34) and (39) into the expression of $\left\{P_2\left(\mathbf{u}^0, \boldsymbol{\alpha}^0; \mathbf{h}_u, \mathbf{h}_\psi, \mathbf{h}_\alpha; \boldsymbol{\psi}^{(1)}, \boldsymbol{\psi}^{(2)}\right)\right\}_{\partial\Omega_x}$ must cause all terms containing unknown values of $\mathbf{h}_u$ and $\mathbf{h}_\psi$ to vanish, which will reduce the bilinear concomitant to an expression that will contain only known values of $\mathbf{h}_\alpha$; we denote this expression as $\hat{P}_2\left(\mathbf{u}^0, \boldsymbol{\alpha}^0; \boldsymbol{\psi}^{(1)}, \boldsymbol{\psi}^{(2)}; \mathbf{h}_\alpha\right)$.

Once the adjoint functions $\boldsymbol{\psi}^{(1)}$ and $\boldsymbol{\psi}^{(2)}$ have been determined by solving Eqs. (38) and (39), they can be used in conjunction with Eq. (37) to represent the "*indirect-effect term*", $(\delta S_i)_{indirect}$, defined in Eq. (27) in the form

$$(\delta S_i)_{indirect} = \left\langle \left(\boldsymbol{\psi}^{(1)}, \boldsymbol{\psi}^{(2)}\right) \begin{pmatrix} \left\langle \mathbf{D}_\alpha Q(\boldsymbol{\alpha}^0) - \mathbf{D}_\alpha\left[\mathbf{L}(\boldsymbol{\alpha}^0)\mathbf{u}^0\right], \mathbf{h}_\alpha \right\rangle_\alpha \\ \left\langle \mathbf{D}_{\alpha u}^2 R(e^0) - \mathbf{L}^*\left[(\boldsymbol{\alpha}^0)\boldsymbol{\psi}^0\right], \mathbf{h}_\alpha \right\rangle_\alpha \end{pmatrix} \right\rangle_{u\psi} - \hat{P}_2\left(\mathbf{u}^0, \boldsymbol{\alpha}^0; \boldsymbol{\psi}^{(1)}, \boldsymbol{\psi}^{(2)}; \mathbf{h}_\alpha\right).$$
(37)

In terms of the adjoint functions $\boldsymbol{\psi}^{(1)}$ and $\boldsymbol{\psi}^{(2)}$, the complete expression of the second-order mixed sensitivities $\delta S_i = (\delta S_i)_{direct} + (\delta S_i)_{indirect}$ is obtained by adding the above expression to the previously computed "direct effect term" from Eq. (26), to obtain

$$\delta S_i\left(\mathbf{u}^0, \boldsymbol{\alpha}^0, \boldsymbol{\psi}^0, \boldsymbol{\psi}^{(1)}, \boldsymbol{\psi}^{(2)}; \mathbf{h}_\alpha\right) = \left(\delta S_i\right)_{direct} + \left(\delta S_i\right)_{indirect}$$

$$= \frac{\partial}{\partial \alpha_i}\left\{\left\langle \mathbf{D}_\alpha R(\mathbf{u},\boldsymbol{\alpha}) - \mathbf{D}_\alpha \hat{P}(\mathbf{u},\boldsymbol{\psi},\boldsymbol{\alpha}) + \left\langle \boldsymbol{\psi}, \mathbf{D}_\alpha \mathbf{Q}(\boldsymbol{\alpha}) - \mathbf{D}_\alpha\left[\mathbf{L}(\boldsymbol{\alpha})\mathbf{u}\right]\right\rangle_\psi, \mathbf{h}_\alpha \right\rangle_\alpha\right\}_{(\mathbf{u}^0,\boldsymbol{\psi}^0;\boldsymbol{\alpha}^0)} \quad (38)$$

$$+ \left\langle \left(\boldsymbol{\psi}^{(1)}, \boldsymbol{\psi}^{(2)}\right) \begin{pmatrix} \left\langle \mathbf{D}_\alpha \mathbf{Q}(\boldsymbol{\alpha}^0) - \mathbf{D}_\alpha\left[\mathbf{L}(\boldsymbol{\alpha}^0)\mathbf{u}^0\right], \mathbf{h}_\alpha\right\rangle_\alpha \\ \left\langle \mathbf{D}^2_{\alpha u}R(e^0) - \mathbf{L}^*\left[(\boldsymbol{\alpha}^0)\boldsymbol{\psi}^0\right], \mathbf{h}_\alpha\right\rangle_\alpha \end{pmatrix}\right\rangle_{u\psi} - \hat{P}_2\left(\mathbf{u}^0, \boldsymbol{\alpha}^0; \boldsymbol{\psi}^{(1)}, \boldsymbol{\psi}^{(2)}; \mathbf{h}_\alpha\right).$$

It is convenient to call Eqs. (38) and (39) *second adjoint sensitivity system* (*SASS*), since, although it itself is not of "second-order", it is paramount for computing the adjoint functions needed for computing the $i^{th}$-row $\left\{\partial^2 R(\mathbf{u},\boldsymbol{\alpha})/\partial \alpha_j \partial \alpha_i\right\}_{(\mathbf{u}^0,\boldsymbol{\alpha}^0)}$, $j=1,...,N_\alpha$, of second-order response sensitivities. Note that the *SASS* is independent of parameter variations $\mathbf{h}_\alpha$. Therefore, these equations need to be solve only once to compute the adjoint functions $\boldsymbol{\psi}^{(1)}$ and $\boldsymbol{\psi}^{(2)}$. Thus, *the exact computation of all second-order sensitivities,* $\left\{\partial^2 R(\mathbf{u},\boldsymbol{\alpha})/\partial \alpha_j \partial \alpha_i\right\}_{(\mathbf{u}^0,\boldsymbol{\alpha}^0)}$, $i,j=1,...,N_\alpha$, *using the SASS requires* $2N_\alpha$ *large-scale (adjoint) computations*, rather that $O(N_\alpha^2)$ large-scale computations as would be required by forward methods. It is also important to note that the construction and solution of the *second adjoint sensitivity system* (*SASS*) requires very little effort beyond that already invested in solving the original forward Eq.(1) for the state variable $\mathbf{u}(\mathbf{x})$ and the adjoint sensitivity Eq. (22) for the adjoint function $\boldsymbol{\psi}$. This is because, as Eq. (38.b) indicates, the equation for determining the adjoint function $\boldsymbol{\psi}^{(2)}$ is the same as that for determining $\mathbf{u}(\mathbf{x})$, except for a different source term, while the equations to be solved for determining the adjoint function $\boldsymbol{\psi}^{(1)}$ is the same as the adjoint equation for determining the adjoint function $\boldsymbol{\psi}$, again save for a different source term.

## 4. CONCLUSIONS

This work has presented the *second-order forward and adjoint sensitivity analysis procedures* (*SO-FSAP and SO-ASAP*) for computing exactly and efficiently the second-order functional derivatives of general (physical, engineering, biological, etc) system responses (i.e., "system performance parameters") to the system's model parameters. The definition of "system parameters" used in this work include, in the most comprehensive sense, all computational input data, correlations, initial and/or boundary conditions, etc. The *SO-ASAP* builds on the first-order adjoint sensitivity analysis procedure (*ASAP*) for nonlinear systems introduced ([1], [2]) and developed ([3]-[5]) by Cacuci; see also [6]. For a physical system comprising $N_\alpha$ parameters and $N_r$ responses, we note the following essential computational properties of the *SO-FSAP and SO-ASAP*, respectively:

(i) The *SO-FSAP* requires $N_\alpha$ large-scale forward model computations to obtain all of the first-order response sensitivities $\{\partial R(\mathbf{u},\boldsymbol{\alpha})/\partial \alpha_i\}_{(\mathbf{u}^0,\boldsymbol{\alpha}^0)}$, $i=1,...,N_\alpha$, and an additional number $N_\alpha(N_\alpha+1)/2$ of large-scale computations to determine the sensitivities $\{\partial^2 R(\mathbf{u},\boldsymbol{\alpha})/\partial \alpha_j \partial \alpha_i\}_{(\mathbf{u}^0,\boldsymbol{\alpha}^0)}$, $i,j=1,...,N_\alpha$, for a total of $(N_\alpha^2/2 + 3N_\alpha/2)$ *large scale computations for obtaining all of the first- and second-order sensitivities, for all $N_r$ system responses*;

(ii) The *SO-ASAP* requires one large-scale adjoint computation for computing all of the first-order response sensitivities $\{\partial R(\mathbf{u},\boldsymbol{\alpha})/\partial \alpha_i\}_{(\mathbf{u}^0,\boldsymbol{\alpha}^0)}$, $i=1,...,N_\alpha$, and an additional number of $2N_\alpha$ large-scale computations to determine the sensitivities $\{\partial^2 R(\mathbf{u},\boldsymbol{\alpha})/\partial \alpha_j \partial \alpha_i\}_{(\mathbf{u}^0,\boldsymbol{\alpha}^0)}$, $i,j=1,...,N_\alpha$, for a *total of* $(2N_\alpha+1)$ *large scale computations for obtaining all of the first- and second-order sensitivities, for one functional-type system responses.*

The above considerations clearly highlight the fact that the *SO-FSAP* should be used when $N_r \gg N_\alpha$, while the *SO-ASAP* should be used when $N_\alpha \gg N_r$, which is the most often encountered situation in practice. The *SO-ASAP* presented, in premiere, in this work should enable the hitherto intractable exact computation of all of the second-order response sensitivities (i.e., functional Gateaux-derivatives) to the large-number of parameters that are characteristically encountered in the large-systems of practical interest. Additional work is currently in progress to generalize the *SO-ASAP* to computing efficiently and exactly the third- and higher-order response sensitivities. In the accompanying PART II [13], we present an illustrative application of the *SO-ASAP* to a paradigm particle diffusion problem that admits a unique analytical solution, thereby making transparent the mathematical derivations presented in this paper. Very importantly, this illustrative application will show that:

(i) The construction and solution of the second adjoint sensitivity system (*SASS*) requires very little additional effort beyond the construction of the adjoint sensitivity system needed for computing the first-order sensitivities; and

(ii) The actual number of adjoint computations needed for computing all of the first- and second-order response sensitivities is considerably less than $2N_\alpha$ per response.

As a final comment, we note that the *SO-ASAP* has been developed in this work using *real* (as opposed to complex) Hilbert spaces; this has been done because real Hilbert spaces provide the natural mathematical setting for computational purposes. This setting does *not* restrict, in any way, the generality of the *AO-ASAP* theory presented here. The *AO-ASAP* theory can be readily set in complex Hilbert spaces by simply changing some terminology, without affecting its substance.

**ACKNOWLEDGMENTS**

This work was partially supported by contract 15540-FC51 from Gen4Energy, Inc., and partially by contract 15540-FC59 from the US Department of Energy (NA-22), respectively, with the University of South Carolina. The author wishes to express his personal appreciation to Mr. Regan Voit, Gen4Energy VP for Applied Research and Manufacturing Initiatives, and Mr. James Peltz, NA-22 Program Manager, for their continued support.